\documentclass[journal]{IEEEtran}
\ifCLASSINFOpdf
  \usepackage[pdftex]{graphicx}
\else
  \usepackage[dvips]{graphicx}
\fi

\ifCLASSOPTIONcompsoc
  \usepackage[caption=false,font=normalsize,labelfont=sf,textfont=sf]{subfig}
\else
  \usepackage[caption=false,font=footnotesize]{subfig}
\fi
\usepackage{comment}
\usepackage{bm}
\usepackage{amsmath}
\usepackage{amsfonts}
\usepackage{xcolor}
\usepackage[normalem]{ulem}
\usepackage{physics}
\usepackage{siunitx}

\usepackage{booktabs}
\usepackage{multicol}
\usepackage{multirow}

\DeclareRobustCommand{\Erase}{\bgroup\markoverwith{\textcolor{red}{\rule[.5ex]{2pt}{0.4pt}}}\ULon}

\begin{document}

%

\title{Tactile Rendering Using Three Basic Stimulus Components in Ultrasound Midair Haptics}

%
%
%
\author{Tao~Morisaki,
        Atsushi~Matsubayashi,
        Yasutoshi~Makino,
        and~Hiroyuki~Shinoda,~\IEEEmembership{Member,~IEEE}
\thanks{Manuscript received xx; revised xx.}
\thanks{This work was supported in part by NEDO SIP 23201554-0 and JSPS KAKENHI 25K21261.}
\thanks{T. Morisaki is with the Communication Science Laboratories, NTT, Inc., Atsugi, Japan. E-mail: tao.morisaki@ntt.com}
\thanks{A. Matsubayashi, Y. Makino, and H. Shinoda are with the Graduate School of Frontier Sciences, the University of Tokyo, Kashiwa-shi, Chiba, 277-8561, Japan (E-mail: matsubayashi@hapis.k.u-tokyo.ac.jp; yasutoshi\_makino@k.u-tokyo.ac.jp; hiroyuki\_shinoda@k.u-tokyo.ac.jp).}
}

%
%

\markboth{xxxx}%
{Shell \MakeLowercase{\textit{et al.}}: Bare Demo of IEEEtran.cls for IEEE Journals}
%



\maketitle


\begin{abstract}
Ultrasound midair haptics (UMH) can present non-contact tactile stimuli using focused ultrasound without any wearables. Recently, UMH has been shown to present not only conventional vibration stimulus but also static pressure stimulus by locally rotating an ultrasound focus at several hertz. Current UMH can present three basic tactile stimuli: static pressure, 30 Hz vibration, and 150 Hz vibration. These primarily elicit responses from three distinct types of mechanoreceptors: SA-I, FA-I, and FA-II. As human texture perception relies on the combination of mechanoreceptor neural responses, this study proposes combining the three basic stimuli to render tactile texture in UMH. Experimental results demonstrate that the proposed method can render at least six discriminable textures with different roughness and friction sensations. Notably, through comparisons with real physical objects, we found that the pressure-only stimulus was perceived as slippery and smooth. The smoothness was similar to a glass marble. When vibration stimuli were synthesized, the perceived roughness and friction increased significantly. The roughness level reached that of a 100-grit sandpaper.
\end{abstract}


%
\IEEEpeerreviewmaketitle

\newcommand{\rOne}{\mathrm{I}}
\newcommand{\rTwo}{\mathrm{I\hspace{-.1em}I}}
\section{Introduction}
Ultrasound midair haptics, which presents a non-contact tactile stimulus without any wearables~\cite{iwamoto2008non, hoshi2010noncontact, carter2013ultrahaptics}, is a promising haptic technology. By focusing ultrasound onto the skin, a non-contact force called acoustic radiation force is generated at the focus~\cite{yosioka1955acoustic}, conveying tactile sensation. Since this method does not require users to wear mechanical devices, pre-determining the stimulation points on the skin is not required. Ultrasound midair haptics has been used for many applications~\cite{rakkolainen2020survey}, including midair interfaces with tactile feedback~\cite{monnai2014haptomime,cornelio2017agency,young2020designing,korres2020mid}, behavioral guidance~\cite{suzuki2019midair,yoshimoto2019midair,freeman2019haptiglow}, and evoking pleasant sensations~\cite{tsumoto2021presentation}.

The reproduction of real object texture has also been actively studied in midair haptics~\cite{freeman2017textured,ablart2019using,sakiyama2020midair,freeman2021enhancing,hosoi2024voluminous}. Sakiyama et al. measured changes in pressure distribution on the skin generated by the stroking of real objects and reproduced them using focused ultrasound~\cite{sakiyama2020midair}. Ablart et al. reported roughness sensations were evoked by rotating an ultrasound focus on the palm, and the roughness sensation was varied when the focus rotation frequency was varied~\cite{ablart2019using}.

\begin{figure}[t]
  \centering
  \includegraphics[width=1\columnwidth]{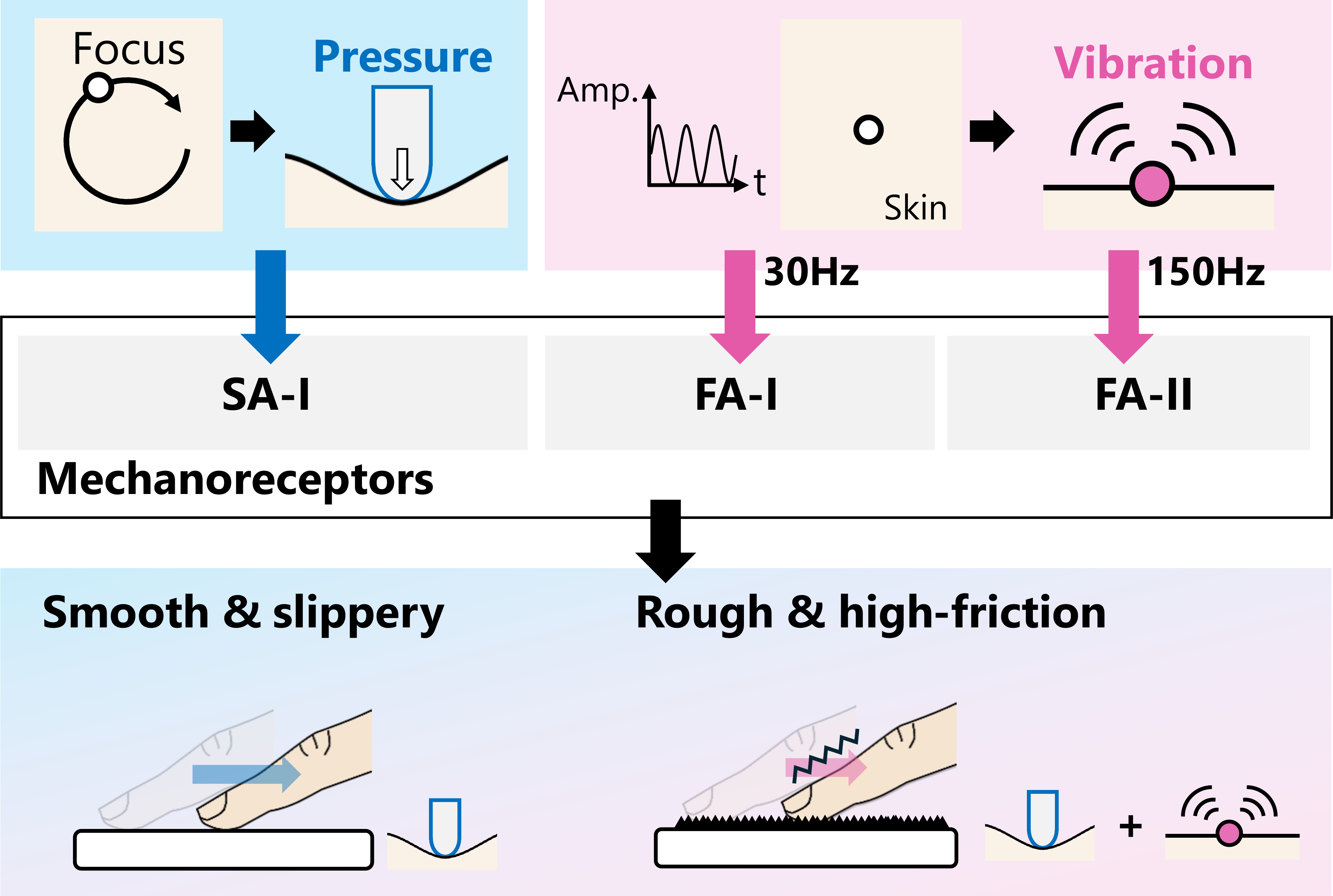}
  \caption{Proposed tactile rendering method, synthesizing three ultrasound stimuli primarily stimulating different mechanoreceptors. For SA-I, a 5 Hz focus rotation, evoking a static pressure sensation, was used. For FA-I and FA-II, vibration stimuli at 30 and 150 Hz were used, respectively. Pressure-only stimulus was perceived as slippery and smooth. When vibration sensations were synthesized, the perceived roughness and friction increased.}
  \label{fig:Fig/Teaser/teaser2.pdf}
\end{figure}

However, perceptual mechanisms-based tactile rendering has remained incomplete in ultrasound midair haptics, as a static pressure stimulus, a basic stimulus component, has been lacking. Humans' texture perception partly relies on the responses of mechanoreceptors with different frequency characteristics: SA-I (Slowly adaptive type-I), FA-I, and FA-II (Fast adaptive type-I and II)~\cite{bolanowski1988four}. FA-I and FA-II primarily respond to vibrations around 30 Hz and 150 Hz, respectively, while SA-I primarily responds to static pressure. Considering this mechanism, synthesizing three types of stimuli corresponding to each receptor is effective for texture reproduction, and this approach has already been adopted in other haptic displays~\cite{konyo2005tactile,yem2017wearable,huang2023skin}. In midair haptics, vibration stimuli around 30 and 150 Hz can be presented by modulating the amplitude of ultrasound. In contrast, static pressure stimulus corresponding to SA-I had not been achieved prior to the study by~\cite{morisaki2021non}, as the radiation force of ultrasound is weak, several tens of millinewtons. Without a pressure stimulus, texture reproduction remains incomplete because skin receptors inevitably detect static pressure during physical contact with a surface. As can be easily inferred, an ideal smooth and slippery surface cannot be rendered in the absence of a pressure stimulus, since vibration sensations evoke roughness and friction~\cite{bensmaia2003vibrations, natsume2017skin,guest2012physics,nonomura2009tactile}.

To address this issue, Morisaki et al. recently demonstrated that moving an ultrasound focus at a few hertz evokes a static pressure sensation~\cite{morisaki2021non,morisaki2023noncontact,morisaki2024towards}. In their experiment, the focus rotated at 5 Hz along a circular trajectory with a radius of several millimetres, and the perceived intensity was equivalent to approximately 0.2 N, ten times stronger than the radiation force~\cite{morisaki2023noncontact}. Because the responses of FA-I and FA-II elicit a vibration sensation, whereas that of SA-I elicits a pressure sensation, the 5 Hz focus rotation is an effective method for primarily stimulating SA-I. 

Building on these studies, we propose a texture rendering method, synthesizing three ultrasound stimuli corresponding to the three receptors, and investigate the renderable texture. For SA-I, we employed a 5 Hz focus rotation to evoke a static pressure sensation. For FA-I and FA-II, we employed vibration stimuli at 30 and 150 Hz, respectively. These vibration stimuli were synthesized with the pressure stimulus by modulating the amplitude of the focus rotating at 5 Hz.

In the experiments, we demonstrated that the proposed method can render at least six discriminable textures with different roughness and friction sensations. The roughness and friction sensations of the ultrasound stimuli were evaluated by comparing them with real physical objects. Notably, when only the pressure stimulus was presented, the resulting sensation was perceived as highly slippery and smooth. Its smoothness level was close to that of a glass marble. When vibration sensations were synthesized, the perceived roughness and friction increased significantly. The roughness level reached that of a 100-grit sandpaper. Although previous studies have presented smooth and rough sensations using focused ultrasound, those approaches did not employ static pressure stimuli and did not conduct a comparison with real materials~\cite{ablart2019using, beattie2020incorporating}. They reported the relative roughness score (magnitude) only.

\section{Related Works}
\subsection{Contact-type Haptic Display}
In contact-type haptic displays, as in our approach, synthesising multiple stimuli to which different mechanoreceptors primarily respond has been proposed to render tactile textures. Konyo et al. simultaneously presented 5 Hz vibration and higher frequency vibrations using a single vibrator~\cite{konyo2005tactile}. The 5 Hz vibration stimulus corresponds to the role of our LM stimulus. They reported that a 5 Hz vibration evoked a static pressure sensation~\cite{konyo2005tactile}. Yem et al. simultaneously presented pressure sensation and low-frequency vibration sensation using electrical stimulation~\cite{yem2017wearable}. They also simultaneously presented higher-frequency vibration sensation and shear force stimulation using a DC motor. Huang et al. developed a thin haptic device with built-in soft electrodes and electromechanical actuators~\cite{huang2023skin}. The electrodes presented pressure sensation and low-frequency vibration, while the actuators presented high-frequency vibration.

These studies employed contact-type haptic displays, whereas our study used non-contact ultrasound stimuli.

\subsection{Ultrasound Midair Haptics}
Our study proposes to synthesize pressure and vibration sensations in ultrasound midair haptics and evaluate its performance on tactile texture rendering. Here, we introduce previous studies on rendering haptic textures using ultrasound. Sakiyama et al. measured the spatiotemporal distribution of pressure when stroking soft real objects on a two-dimensional pressure sensor and reproduced the pressure distribution using focused ultrasound~\cite{sakiyama2020midair}. Hosoi et al. rendered voluminous fur sensation using STM stimulus~\cite{hosoi2024voluminous}. They modeled the pressure pattern when stroking fur, and adjusted the movement frequency and intensity of the STM based on the model. Ablart et al. reported that in STM stimuli, in which an ultrasound focus rotates palm-sized, the roughness sensation was varied, related to the movement frequency and trajectory length~\cite{ablart2019using}. Using the STM method, Beattie et al. generated ultrasound roughness stimuli from visual texture images~\cite{beattie2020incorporating}. Freeman showed that the roughness of the STM was enhanced by modulating the amplitude of the rotating focus with white noise~\cite{freeman2021enhancing}. Freeman et al. rendered macro-scale roughness textures by dynamically changing focus position~\cite{freeman2017textured}. With their method, a focus position was moved between a virtual surface, such as an arranged convex surface, and the user's hand. Multi-modal approach was also proposed~\cite{montano2023sounds,xue2024fabsound}. Montano et al. evaluated the effect of auditory stimulus on perception of texture rendered by ultrasound~\cite{montano2023sounds}. Xue et al. rendered fabric texture simultaneously using ultrasound tactile stimuli and auditory stimuli and evaluated their roughness~\cite{xue2024fabsound}. In addition to these rendering specific haptic sensations, general stimulus design frameworks have also been proposed~\cite{mulot2021dolphin,seifi2023feellustrator,lim2024designing}. 

These studies did not use pressure sensation to render haptic texture. This paper is the first to synthesize pressure and vibration sensations to render textures with different roughness and friction sensations using ultrasound stimuli. As an exception, Morisaki et al. proposed a concept of synthesizing pressure and vibration sensations~\cite{morisaki2021midair}; however, they did not evaluate its performance.

\section{Experimental Setup}

\subsection{Overview\label{sec: Setup Overview}}
\newcommand{\xC}{x^{c}}
\newcommand{\yC}{y^{c}}
\newcommand{\zC}{z^{c}}
\newcommand{\pC}{p^{c}}
\newcommand{\yHand}{y^{hand}}

\begin{figure}[t]
  \centering
  \includegraphics[width=1\columnwidth]{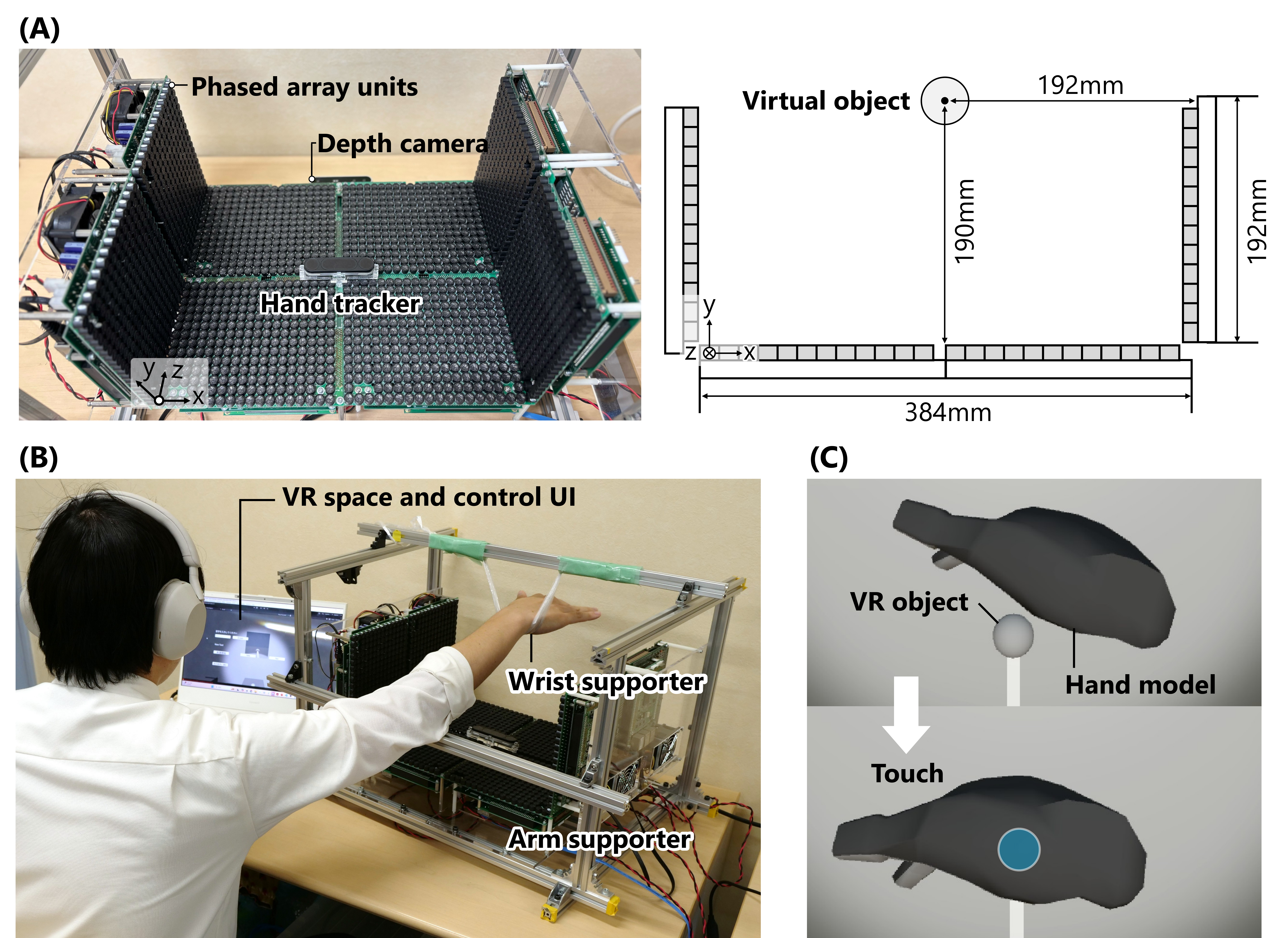}
  \caption{Ultrasound tactile display system we developed. This system was used in Experiment 1--4. (A) Photograph and schematic illustration. (B) A photo of a participant using the system. (C) Screenshot of the VR object and hand model. When a hand model touched a spherical VR object, a tactile stimulus was presented at the contact point on the real hand.}
  \label{fig:Fig/Setup/System.png}
\end{figure}

This ultrasound tactile display system shown in Fig.~\ref{fig:Fig/Setup/System.png}A was used in all experiments conducted in this study. Fig.~\ref{fig:Fig/Setup/System.png}B and C shows an overview of the provided tactile experience. The user’s 3D hand model and a spherical virtual object were presented on a PC display. The position of spherical virtual object was (192, 190, 151) mm. Users can control the hand model by moving their real hand. When the hand model touched the virtual object, a tactile stimulus was presented to the user’s real hand.

The tactile display system consists of eight ultrasound phased array units~\cite{suzuki2021autd3}, a hand tracker (Leap Motion 2, Ultraleap, Ltd.), a depth camera (RealSense D415, Intel), and a control PC. The coordinate system of the tactile display system is shown in Fig.~\ref{fig:Fig/Setup/System.png}. 

An ultrasound phased array, an array of individually controllable ultrasound transducers, was used to present tactile stimuli. A phased array can generate an ultrasound focus at arbitrary three-dimensional positions by controlling the phase of each transducer~\cite{hoshi2010noncontact, carter2013ultrahaptics}. A total of 1,992 transducers were used, with an emitted ultrasound frequency of 40 kHz~\cite{suzuki2021autd3}. 

The hand tracker and the depth camera were used to determine the timing and position for presenting ultrasound stimuli. The depth camera was used to present the ultrasound focus accurately on the surface of the hand using the obtained depth data. First, the hand tracker detected the participant's hand and generated a hand model. When the hand model touched a spherical VR object, the y-position of the real hand $\yHand$ at the contact position was measured with the depth camera. Based on the detected y-position, the center position of the ultrasound stimulus $\pC$ was determined as follows:
\begin{IEEEeqnarray}{rCl}
    \pC = (\xC, \yHand, \zC),
\end{IEEEeqnarray}
where $\xC$ and $\zC$ are the x and z positions of the spherical VR objects' center. This measurement process ran at 90 Hz.

\subsection{Stimulus Design}
\newcommand{\xLM}{x^{LM}}
\newcommand{\yLM}{y^{LM}}
\newcommand{\zLM}{z^{LM}}
\newcommand{\fLM}{f^{LM}}
\newcommand{\pLM}{\boldsymbol{p}^{LM}}
\newcommand{\rLM}{r}
\newcommand{\dt}{dt}

\newcommand{\SoundPre}{P}
\newcommand{\SoundPreMax}{P^{max}}
\newcommand{\Amp}{A}
\newcommand{\AmpAM}[1]{A^{AM}_{#1}}

\newcommand{\AmpMax}{A^{max}}
\newcommand{\fAM}[1]{f^{AM}_{#1}}

\begin{figure*}[t]
  \centering
  \includegraphics[width=2\columnwidth]{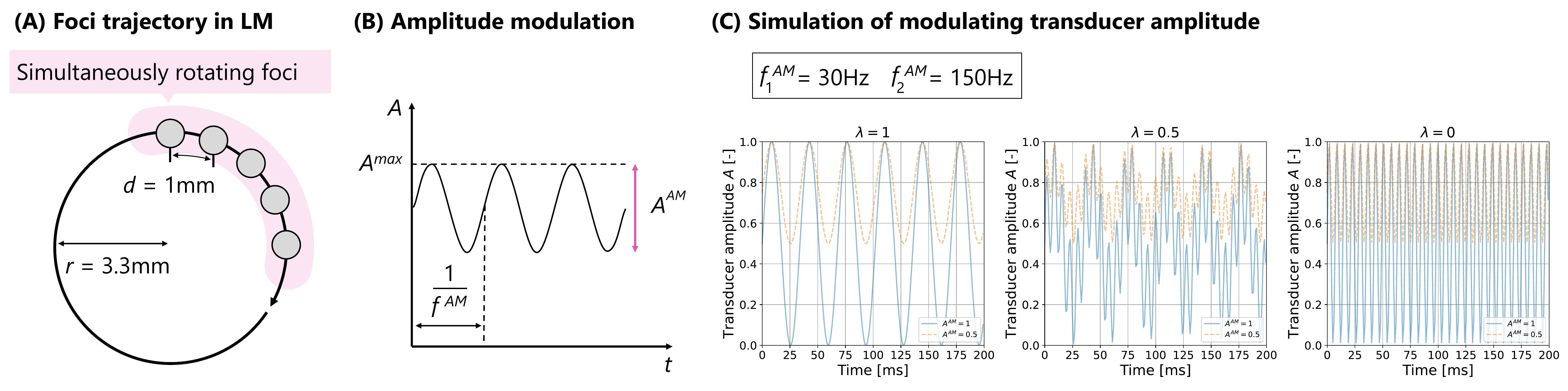}
  \caption{Schematic illustration of ultrasound stimulation design. (A) LM in which 5 foci rotate at 5 Hz. (B) Variation of sound amplitude $\Amp$ with respect to modulation amplitude $\AmpAM{}$. (C) Simulated modulated transducer amplitude $\Amp$. $\fAM{1} = 30$ Hz and $\fAM{2} = 150$ Hz were used for this simulation.}
  \label{fig:Fig/Setup/SchematicStimulus.pdf}
\end{figure*}

This section describes the stimulus design to synthesize pressure and vibration sensations. A schematic of the stimulus design is shown in Fig.~\ref{fig:Fig/Setup/SchematicStimulus.pdf}A and B. First, to present a pressure sensation, we used an LM stimulus~\cite{morisaki2021non}. In the LM, five ultrasound foci were simultaneously presented and rotated since such multiple foci rotation is more suitable for producing static pressure sensation~\cite{morisaki2023noncontact}. The previous study reported that a non-static sensation, such as a perception of focused movement, was suppressed using a multi-foci condition compared with a single focus rotation condition~\cite{morisaki2023noncontact}. The multiple foci rotation has also been used in STM, but these studies did not aim to present static pressure sensations~\cite{shen2023multi,shen2024controlled}. To synthesize vibration sensation with the pressure sensation, the amplitude of these five foci was modulated with sinusoidal waves of 30 and 150 Hz. We also used a composite wave of 30 and 150 Hz waves as a modulation waveform. Hajas et al. have used a similar method, amplitude-modulated STM stimulus; however, they aimed to render 2D tactile shapes and not to simultaneously present pressure and vibration sensations~\cite{hajas2020mid}.  

\subsubsection{Formulation}
We first formulate the five foci trajectory of LM $\pLM_i$. The $i \in \{1,...5\}$ is the focus number simultaneously presented in the LM. The $\pLM_i(t) = (\xLM_i, \yLM_i, \zLM_i)$ is as follows: 
\begin{IEEEeqnarray}{rCl}
\xLM_i &=& \xC + \rLM\cos(2\pi\fLM t + \phi^{LM}_{i}),\\
\yLM_i &=& \yHand,\\
\zLM_i &=& \zC + \rLM\sin(2\pi\fLM t + \phi^{LM}_{i}),\\
\phi^{LM}_{i} &=& \frac{d}{\rLM}(i-1) 
\end{IEEEeqnarray}
where $\fLM$ [Hz] is the focus rotation frequency, $\rLM$ [mm] is the radius of the focus trajectory, and $d$ [mm] is the space of simultaneously presented multiple foci. The foci positions were updated at 1000 Hz. The driving signal of each transducer was calculated using a linear synthesis scheme as in the previous study~\cite{morisaki2023noncontact}. 

Second, we formulate the modulation of transducer amplitude for presenting vibration sensation. All transducers were driven with the same amplitude. For modulation, the transducer amplitude $\Amp \in [0, 1]$ was formulated as follows:
\begin{IEEEeqnarray}{rCl}
\Amp &= \Amp^{max}\Amp^{mod},
\end{IEEEeqnarray}
where $\Amp^{max} \in [0, 1]$ is the coefficient to manipulate final transducer amplitude. If $\AmpMax = 0$, ultrasound is not output. If $\AmpMax = 1$, the ultrasound is output with the maximum sound pressure. $\Amp^{mod} \in [0, 1]$ is the coefficient to modulate the amplitude to present vibration sensations. $\Amp^{mod}$ is formulated as follows:
\begin{IEEEeqnarray}{rCl}
\Amp^{mod} &=& \AmpAM{}(\lambda\Phi_1 + (1 - \lambda)\Phi_2 )+ 1 - \AmpAM{},\label{eq:AM}\\
\Phi_1 &=& \frac{1}{2}(\cos{(2\pi\fAM{1}t)}+1),\\
\Phi_2 &=& \frac{1}{2}(\cos{(2\pi\fAM{2}t)}+1),
\end{IEEEeqnarray}
where $t$ is the elapsed time. $\AmpAM{} \in [0,1]$ is the modulation amplitude. With $\AmpAM{} = 0$, only the LM stimulus with the constant amplitude was presented. $(\fAM{1}, \fAM{2}) = (30, 150)$ Hz is the modulation frequency. $\lambda \in [0,1]$ is the allocation ratio of the modulation amplitude between the modulation frequencies. If $\lambda = 1$, the amplitude was modulated only at $\fAM{1}$. If $\lambda = 0$, the amplitude was modulated only at $\fAM{2}$. If $\lambda = 0.5$, the amplitude was modulated at both $\fAM{1}$ and $\fAM{2}$ with the same level. The simulation example of transducer amplitude $\Amp$ with eq.~\ref{eq:AM} is shown in Fig.~\ref{fig:Fig/Setup/SchematicStimulus.pdf}C. In the simulation, ($\fAM{1}, \fAM{2}) = (30, 150)$ Hz, $\AmpAM{} = 0.5, 1$, and $\lambda = 0, 0.5, 1$ were used.

We expected that the intensity of vibration sensation would increase with respect to the $\AmpAM{}$, and confirmed this expectation in Experiments 1 and 2.

\subsubsection{Stimuli Used in Experiments\label{sec:Designed stimuli used in experiments}}
\begin{table}[t]
  \centering
  \caption{Summary of stimuli used in the experiments.}
  \begin{tabular}{@{}ll@{}}
    \toprule
    Stimulus & $\lambda$\\
    \midrule
    \textit{S-30Hz}   & 1  \\
    \textit{S-150Hz}  & 0  \\
    \textit{S-Mix1}   & 0.5\\
    \textit{S-Mix2}   & 0.7\\
    \textit{S-LM}     & - \\
    \bottomrule
\end{tabular}
\label{table:Summary of stimuli used in the experiments}
\end{table}

We designed five stimuli by manipulating $\lambda$ in eq.~\ref{eq:AM}. The designed stimuli are summarized in Table~\ref{table:Summary of stimuli used in the experiments}.  

First, we designed an LM stimulus evoking pressure sensation only~\cite{morisaki2021non, morisaki2023noncontact}. The condition was $\fLM = 5$ Hz,  $d = 1$ mm, and $\rLM = 3.3$ mm. This LM stimulus without amplitude modulation was referred to as \textit{S-LM}. 

Second, for synthesizing vibration sensation, we designed four amplitude-modulated LM stimuli using eq.~\ref{eq:AM}: \textit{S-30Hz}, \textit{S-150Hz}, \textit{S-Mix1}, and \textit{S-Mix2}. For all stimuli, the modulation frequency $(\fAM{1}, \fAM{2}) = (30, 150)$ Hz was used. The $\lambda = 1$ and $\lambda = 0$ were used for S-30Hz and S-150Hz, respectively. The $\lambda = 0.5$ and $\lambda = 0.7$ were also used for S-Mix1 and S-Mix2, respectively. The S-Mix2, in which the physical intensity of the 30 Hz vibration was higher than that of the 150 Hz vibration, was designed because the 30 Hz vibration was perceived as weaker than the 150 Hz vibration in our preliminary experiment. The modulation amplitude $\AmpAM{}$ was varied from 1 to 0 in Experiments 1 and 2. In Experiments 3 and 4, different $\AmpAM{}$ was used for each stimulus. The different $\AmpAM{}$ design is as described in Section~\ref{sec:Ex3 Setup and Stimulus}.  

We conducted sound pressure measurement and confirmed that the desired stimulus was presented. The measurement setup and results were described in the Supplemental file.

\section{Exp. 1: Perception Threshold of Vibration}
This experiment evaluated the threshold of modulation amplitude $\AmpAM{}$ for vibration perception to investigate whether the proposed method can synthesize vibration sensations to pressure sensations.  

\subsection{Setup and Stimulus}
We used the tactile system shown in Fig.~\ref{fig:Fig/Setup/System.png}. Participants operated this system while looking at the PC display. The PC display presented both the virtual hand model of the participant’s hand and a spherical VR object. A user interface (UI) for operating was also presented. Participants touched the VR object with the center of their right palm using a wrist and arm supporter as shown in Fig.~\ref{fig:Fig/Setup/System.png}B.

Four ultrasound tactile stimuli, S-30Hz, S-150Hz, S-Mix1, and S-Mix2, as described in Section~\ref{sec:Designed stimuli used in experiments}, were presented.

\subsection{Procedure}
14 males (12 in their 20s and 2 in their 30s) and 4 females (all in their 20s) participated in the experiment. All experiments conducted in this study were approved by the ethics committee at The University of Tokyo (Approval number: 24-626).

We used an interleaved staircase method to manipulate the modulation amplitude $\AmpAM{}$, and the $\AmpAM{}$ varied from 0 to 1 by 0.02 according to the participant's response. Before the experiment, participants experienced the five stimuli: S-LM, S-30Hz, S-150Hz, S-Mix1, and S-Mix2. An ultrasound tactile stimulus was presented to the participant’s palm for 1 s. After this stimulus, participants responded whether they perceived vibration with “yes” or “no”. The stimulus presentation and the response were conducted by clicking buttons displayed on the PC display. In the ascending series, an initial $\AmpAM{}$ was 0, corresponding to S-LM evoking pressure sensation only~\cite{morisaki2021non,morisaki2023noncontact}. If the response was “no” in this ascending series, the $\AmpAM{}$ increased by 0.02 and the stimulus was presented again. This increase in the $\AmpAM{}$ was repeated until the response reversed, i.e., until the participant perceived vibration. After the reversal, the $\AmpAM{}$ started to decrease by 0.02. This decrease was repeated until the response reversed again. These experimental trials were repeated until the response reversed six times, and the $\AmpAM{}$ at each reversal was recorded. In the descending series, an initial $\AmpAM{}$ was the maximum value of 1, and the $\AmpAM{}$ was decreased by 0.02 until the response reversed. After the reversal, $\AmpAM{}$ started to increase until the response reversed again. The experimental trial was repeated until the response reversed six times, and $\AmpAM{}$ at each reversal was recorded. The ascending and descending series were used in random order, and 12 $\AmpAM{}$ at the response reversal was obtained. The average of the last three reversal $\AmpAM{}$ values obtained in the ascending and descending series was calculated, and it was taken as the perceptual threshold. After that, the stimulus condition was changed. The four stimulus conditions were used in random order. 

\subsection{Results and Analysis}
\begin{figure}[t]
  \centering
  \includegraphics[width=0.7\columnwidth]{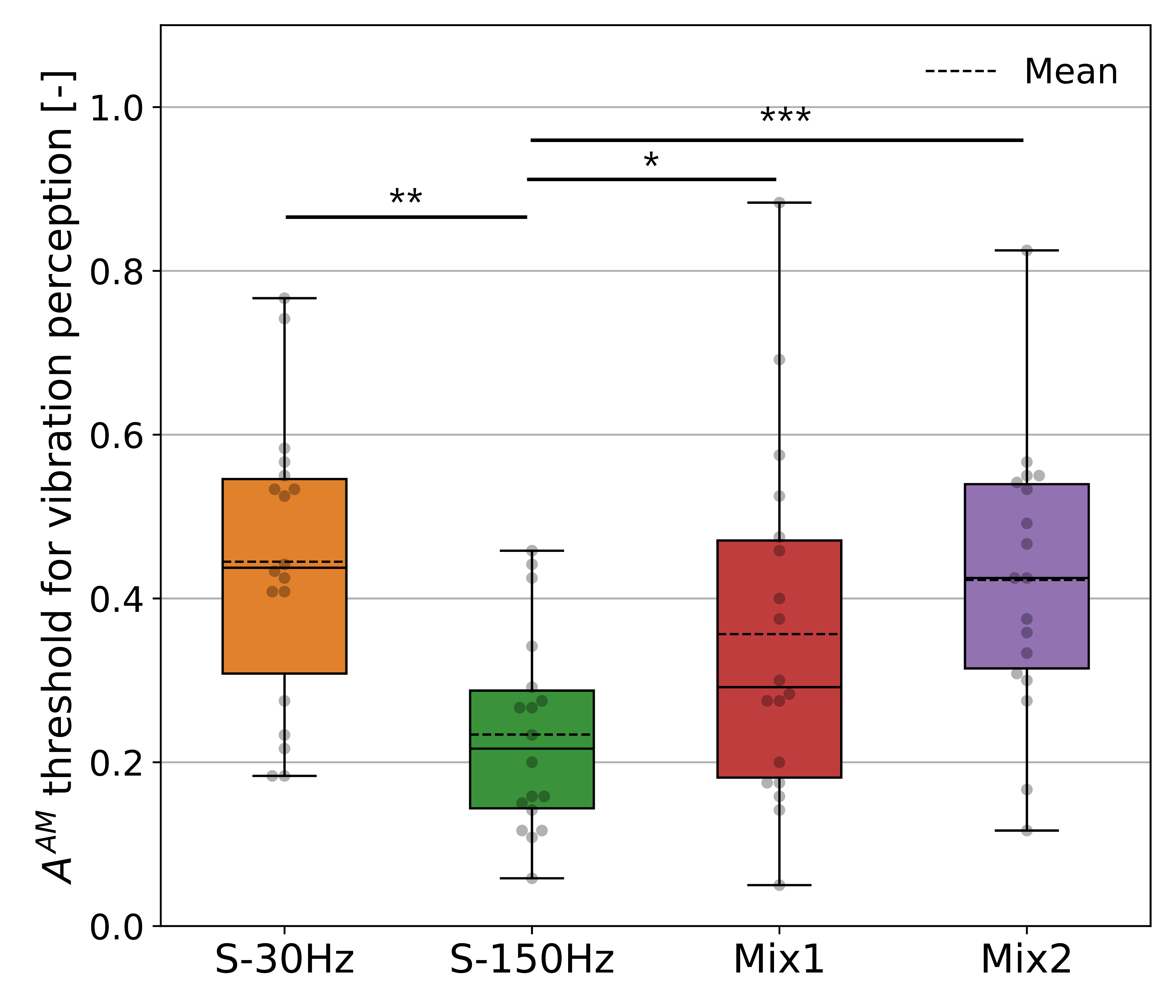}
  \caption{Result of Experiment 1. Threshold modulation amplitude $\AmpAM{}$ for vibration perception. These results indicate that participants were most sensitive to the vibration sensation in S-150Hz.}
  \label{fig:Fig/Ex1/Threshold.pdf}
\end{figure}

Fig.~\ref{fig:Fig/Ex1/Threshold.pdf} shows a boxplot of the thresholds of $\AmpAM{}$ for vibration perception. The stimulus condition with the lowest perceptual threshold was S-150Hz, with an average of 0.23.

We conducted statistical analysis to evaluate the differences in the thresholds across the stimulus conditions. Since the results of the Shapiro-Wilk test indicated that all threshold data were normally distributed ($p > 0.05$), we used parametric analysis. The one-way repeated-measures ANOVA showed that the stimulus condition had a significant effect on perceptual threshold ($F(3, 51) = 8.82, p = 0.000083, {\eta_p}^{2} = 0.34$). As a post-hoc test, we conducted a paired t-test with Holm correction. The results are shown in Fig.~\ref{fig:Fig/Ex1/Threshold.pdf}. The perceptual threshold of S-150Hz was significantly lower than that of all other stimulus conditions ($p < 0.05$). The p-value and Cohen's $d$ were $(p = 0.0026, d= 1.43)$, $(p = 0.00041, d = 1.33)$, and $(p = 0.019, d = 0.72)$ for [S-150Hz, S-30Hz], [S-150Hz, S-Mix2], and [S-150Hz, S-Mix1].

\subsection{Discussion}
These results indicated that vibration sensation was perceived by modulating the amplitude of the rotating ultrasound foci. 

The characteristic of the threshold was consistent with common vibration stimulus~\cite{vallbo1984properties}. The threshold of S-150Hz was significantly lower than that of S-30Hz. A similar tendency has been observed in the common vibration generated using a vibrator.

The results of S-Mix1 and S-Mix2 indicated that adding a 30 Hz component increased the threshold of 150 Hz vibration. The perceptual threshold of S-Mix1 and S-Mix2 was significantly higher than that of S-150Hz.

\section{Exp. 2: Perceived Intensity}
This experiment evaluated the perceived intensities of pressure and vibration sensations with respect to the variation of the modulation amplitude $\AmpAM{}$.

\subsection{Setup and Stimulus}
The setup for this experiment was the same as in Experiment 1. S-30Hz, S-150Hz, S-Mix1, and S-Mix2 as described in Section~\ref{sec:Designed stimuli used in experiments}, were presented. with the maximum $\AmpMax$.

\subsection{Procedure}
14 males (12 in their 20s and 2 in their 30s) and 4 females (all in their 20s) participated in the experiment. 

The ultrasound tactile stimuli with different $\AmpAM{}$ were presented, and participants evaluated the vibration sensation intensity included in them. The evaluation was conducted with a 0--100 point visual analog scale (VAS). First, as a reference, a tactile stimulus with $\AmpAM{}$ = 1 was presented to the palm for 1 s. Next, as a comparison stimulus, a tactile stimulus with a manipulated $\AmpAM{}$ was presented for 1 s. The modulation type of the reference stimulus and the comparison stimulus was the same. To make participants focus only on the vibration sensation, the foci presented in the reference stimulus did not rotate. This $\AmpAM{}$ was manipulated from 0 to 1 by 0.1 (in 11 steps) and used in random order. After experiencing these stimuli, participants evaluated the intensity of vibration sensation perceived in the comparison stimulus on a 0--100 point scale using a slider. A score of 100 corresponded to the vibration intensity of the reference stimulus, i.e., they reported 100 points if the perceived vibration intensity of the comparison stimulus was the same as that of the reference. If they perceived only pressure sensation without vibration, they reported 0 points. After using all steps of $\AmpAM{}$, the stimulus type was changed. The order of the stimulus type was random. Each participant reported the intensity of vibration sensation 44 times (4 stimulus types $\times$ 11 steps of $\AmpAM{}$ = 44 times).

After the vibration evaluation, participants evaluated the intensity of pressure sensation using a 0--100 point VAS. The S-LM was presented to the participant’s palm for 1 s as a reference. The comparison stimuli were the same as those used for the evaluation of vibration sensation intensity. The $\AmpAM{}$ was manipulated from 0 to 1 by 0.1 (in 11 steps). Participants reported the intensity of pressure sensation in the comparison stimulus on a 0--100 point scale using a slider. They were instructed to treat the pressure sensation intensity of the reference as 100. Each participant evaluated the intensity of pressure sensation 44 times (4 stimulus types $\times$ 11 steps of $\AmpAM{}$ = 44 times.)

\subsection{Results and Analysis}
\begin{figure*}[t]
  \centering
  \includegraphics[width=1.9\columnwidth]{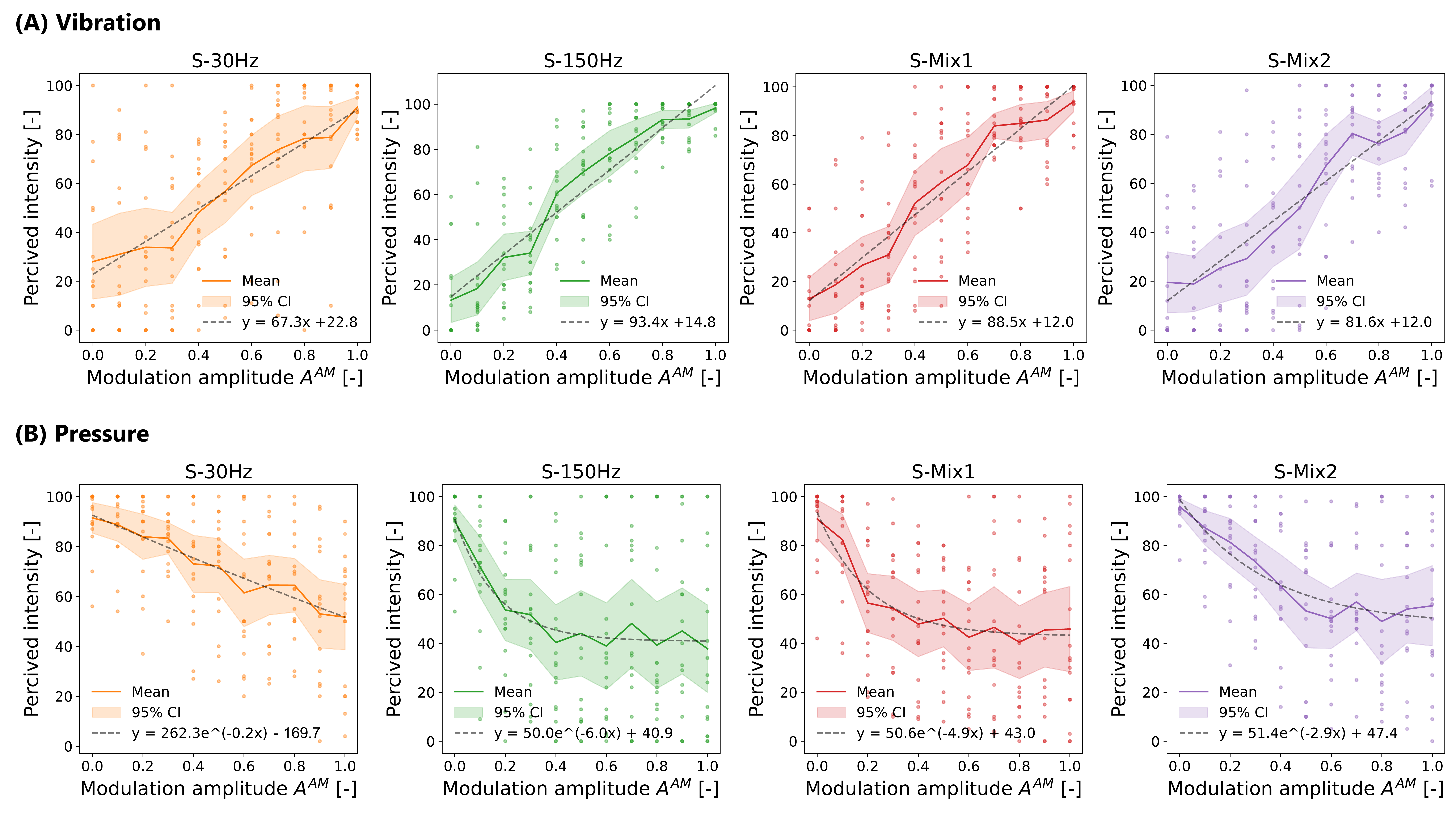}
  \caption{Results of Experiment 2. (A) shows perceived intensity of vibration, and (B) shows pressure sensations for modulation amplitude $\AmpAM{}$. These results indicate that the vibration sensation intensity of S-150Hz increased most rapidly, while its pressure sensation intensity decreased most rapidly.}
  \label{fig:Fig/Ex2/Intensity.pdf}
\end{figure*}

Fig.~\ref{fig:Fig/Ex2/Intensity.pdf}A and B show the intensity variation of vibration and pressure sensation with respect to $\AmpAM{}$.

A Shapiro-Wilk test showed that the data at 43 out of 80 conditions were not normally distributed; thus, we applied the Friedman test to the data with the factor of $\AmpAM{}$. $\AmpAM{}$ has a significant effect on the pressure and vibration intensity at all stimulus conditions. In vibration intensity, the p-values were $p= 2.2\times 10^{-19}, 4.2\times 10^{-27}, 7.6\times 10^{-25}, 4.1\times 10^{-23}$ for S-30Hz, S-150Hz, S-Mix1, and S-Mix2, respectively. In pressure intensity, the p-values were $p= 1.7\times 10^{-12}, 2.9\times 10^{-7}, 3.1\times 10^{-9}, 5.2\times 10^{-11}$ for S-30Hz, S-150Hz, S-Mix1, and S-Mix2, respectively.  

These results indicated that as $\AmpAM{}$ increased, vibration sensation intensity significantly increased while pressure sensation intensity significantly decreased.

We quantitatively evaluated the trends of vibration and pressure sensation intensities by fitting functions. First, a linear function was fitted to the vibration data. The fitted function is as follows:
\begin{IEEEeqnarray}{rCl}
    I = a\AmpAM{} + b,
\end{IEEEeqnarray}
where $I$ is the perceived intensity. Second, an exponential function was fitted to the pressure data since the averaged pressure sensation intensity decreased with the increase of $\AmpAM{}$, and the intensity stopped decreasing around 40 in S-150Hz, S-Mix1, and S-Mix2. The fitted function is as follows:
\begin{IEEEeqnarray}{rCl}
    I = c\mathrm{e}^{a\AmpAM{}} + b.
\end{IEEEeqnarray}
These fitted results are shown in Fig.~\ref{fig:Fig/Ex2/Intensity.pdf}. The $a$, coefficient of the $\AmpAM{}$. All $R^2$ values were higher than 0.9, indicating the functions were well fitted. 

S-150Hz had the largest increase rate of the vibration sensation of $a = 93.4$, and S-30Hz had the lowest of $a=67.3$. S-150Hz also had the largest decrease rate of the pressure sensation $a=-6$, and S-30Hz had the lowest $a=-0.17$.

\subsection{Discussion}
These results showed that the proposed method presents pressure and vibration sensations simultaneously, and that the intensity of vibration sensation can be controlled by $\AmpAM{}$. 

The pressure sensation intensity decreased when the vibration sensation intensity increased, indicating a trade-off between the two sensations. This trade-off is attributed to the acoustic power. When modulating sound pressure, the 5 Hz component of the acoustic power evoking pressure sensation was distributed to the 30 Hz and 150 Hz components, which may result in a decrease in pressure sensation intensity.

The slower decrease in pressure sensation intensity with S-30Hz may be due to the frequency characteristics of Merkel cells, the mechanoreceptors evoking pressure sensation. Since Merkel cells are more sensitive to 30 Hz than to 150 Hz~\cite{bolanowski1988four}, the increased 30 Hz power may have contributed to maintaining pressure sensation intensity.

Furthermore, these results are consistent with the perceptual threshold results obtained in Experiment 1 (shown in Fig.~\ref{fig:Fig/Ex1/Threshold.pdf}). Experiment 1 showed the perceptual threshold of $\AmpAM{}$ in S-150Hz was the lowest, and Experiment 2 showed that the intensity of vibration sensation for S-150Hz most rapidly increased.

\section{Exp. 3: Discriminable Stimulus}
This experiment investigated how many types of discriminable stimuli can be created by the proposed tactile rendering method.

\subsection{Setup and Stimulus\label{sec:Ex3 Setup and Stimulus}}
\begin{table}[t]
  \centering
  \caption{Summary of stimuli used in Experiments 3 and 4.}
  \begin{tabular}{@{}ll@{}}
    \toprule
    Stimulus          & $\AmpAM{}$\\
    \midrule
    \textit{S-LM}       & 0 \\
    \textit{S-30Hz-w}   & 0.5 \\
    \textit{S-30Hz-s}   & 1 \\
    \textit{S-150Hz-w}  & 0.3 \\
    \textit{S-150Hz-s}  & 1 \\
    \textit{S-Mix2}     & 1 \\
    \bottomrule
\end{tabular}
\label{table: Modulation amplitude of stimuli used in Experiment 3 and 4}
\end{table}

The setup for this experiment was the same as in Experiment 1 (shown in Fig.~\ref{fig:Fig/Setup/System.png}). 

We designed six stimuli based on S-LM, S-30Hz, S-150Hz, and S-Mix2. To design stimuli with weak vibration sensation, we manipulated $\AmpAM{}$. As a strong vibration condition, $\AmpAM{} = 1$ was used for S-30Hz and S-150Hz. They were referred to as \textit{S-30Hz-s} and \textit{S-150Hz-s} (s: strong vibration). As a weak vibration condition, $\AmpAM{} = 0.5$ and $\AmpAM{} = 0.3$ were used for S-30Hz and S-150Hz. They were referred to as \textit{S-30Hz-w} and \textit{S-150Hz-w} (w: weak vibration). Because perceived vibration intensity at 150 Hz is higher than that at 30 Hz, as shown in Fig.~\ref{fig:Fig/Ex1/Threshold.pdf} and ~\ref{fig:Fig/Ex2/Intensity.pdf}, we used $\AmpAM{}$ for S-30Hz that was larger than $\AmpAM{}$ for S-150Hz. For S-Mix2, only $\AmpAM{} = 1$ was used. We only used S-Mix2, not S-Mix1, because we focused on stimuli with vastly different tactile sensations. In our preliminary study, the tactile sensation of S-Mix1 was similar to S-150Hz. We used six stimuli: S-LM, S-30Hz-w, S-30Hz-s, S-150Hz-w, S-150Hz-s, and S-Mix2, summarized in Table~\ref{table: Modulation amplitude of stimuli used in Experiment 3 and 4}.

\subsection{Procedure}
10 males (8 in their 20s and 2 in their 30s) and 2 females (all in their 20s) participated in the experiment. 

Before the discrimination test, the maximum sound amplitude $\AmpMax$ was adjusted so that the perceived intensity was equal across stimuli for fair evaluation. In this adjustment phase, the reference stimulus was presented on the participant's palm, and then the comparison stimulus was presented. Participants adjusted the $\AmpMax$ of the comparison stimulus using a slider so that the perceived intensity of the comparison stimulus matched that of the reference. They can experience the standard and comparison stimuli as many times as they wanted. The reference was S-30Hz-s with $\AmpMax = 0.9$. This condition was empirically employed because it was perceived as the weakest in our preliminary tests. The comparison stimuli were the other five stimuli (S-LM, S-30Hz-w, S-150Hz-w, S-150Hz-s, and S-Mix2), and they were presented in random order. In the subsequent discrimination phase, this adjusted $\AmpMax$ was used for each condition. For S-30Hz-s, $\AmpMax = 0.9$ was used.

In the discrimination phase, participants first clicked the “Reference Stimulus” button on the UI, and then they experienced the reference stimulus. As a reference, the six stimuli (as described in Section~\ref{sec:Ex3 Setup and Stimulus}) were presented randomly. Next, participants clicked one of the buttons numbered 1–6 on the UI, and the participants experienced the comparison stimulus. The six stimuli were randomly assigned to the six buttons. After experiencing the stimuli, participants tried to find the stimulus identical to the reference among the six comparison stimuli. They can experience the reference and comparison stimuli as many times as they wanted. Each stimulus was presented for 0.5 s. The stimulus duration was set shorter than in Experiments 1 and 2 so that the participants can quickly switch stimuli many times. Each stimulus was presented twice in random order as the reference. Thus, participants conducted 12 discrimination trials (6 stimuli $\times$ 2 repetitions = 12).  

\subsection{Result}
\begin{figure}[t]
  \centering
  \includegraphics[width=0.8\columnwidth]{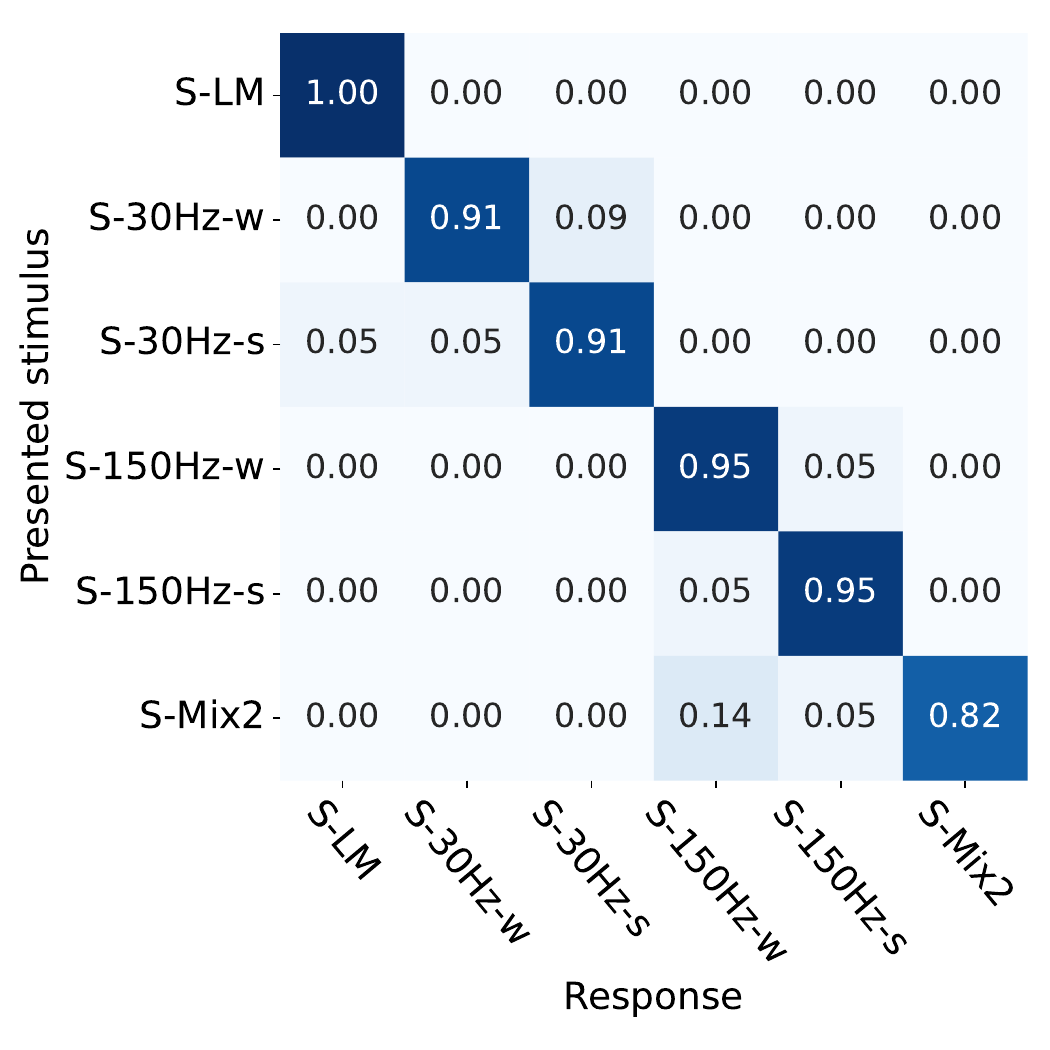}
  \caption{Result of Experiment 3. Discrimination accuracy of the ultrasound stimulus.}
  \label{fig:Fig/Ex3/Accuracy.pdf}
\end{figure}

Fig.~\ref{fig:Fig/Ex3/Accuracy.pdf} shows the confusion matrix, showing the accuracy of the discriminated test. The accuracy of all stimulus conditions exceeded the chance level of 0.167. The highest discrimination rate was 1.0, and the condition was S-LM. The lowest was 0.82, and the condition was S-Mix2.

\subsection{Discussion}
These results indicated that by synthesizing pressure and vibration sensations, at least six discriminable stimuli can be created. Even when the synthesized vibration intensity was not maximum ($\AmpAM{} < 1$), the tactile sensation was still clearly discriminated. These findings indicated that the proposed method is effective in extending the displayable sensation in midair ultrasound haptics. 

Furthermore, participants can discriminate S-Mix2 from other tactile stimuli, including single-frequency vibration sensations (i.e., S-30Hz and S-150Hz). This suggests that amplitude modulation with multiple frequencies can further extend the range of ultrasound tactile stimuli. 

\section{Exp. 4: Evaluating Texture Sensation}
This experiment evaluates the sensation of roughness, stiffness, and friction sensations of the vibration-pressure-synthesized tactile stimuli by comparing them with real materials.  

\subsection{Setup and Stimulus}
\begin{figure}[t]
  \centering
  \includegraphics[width=0.9\columnwidth]{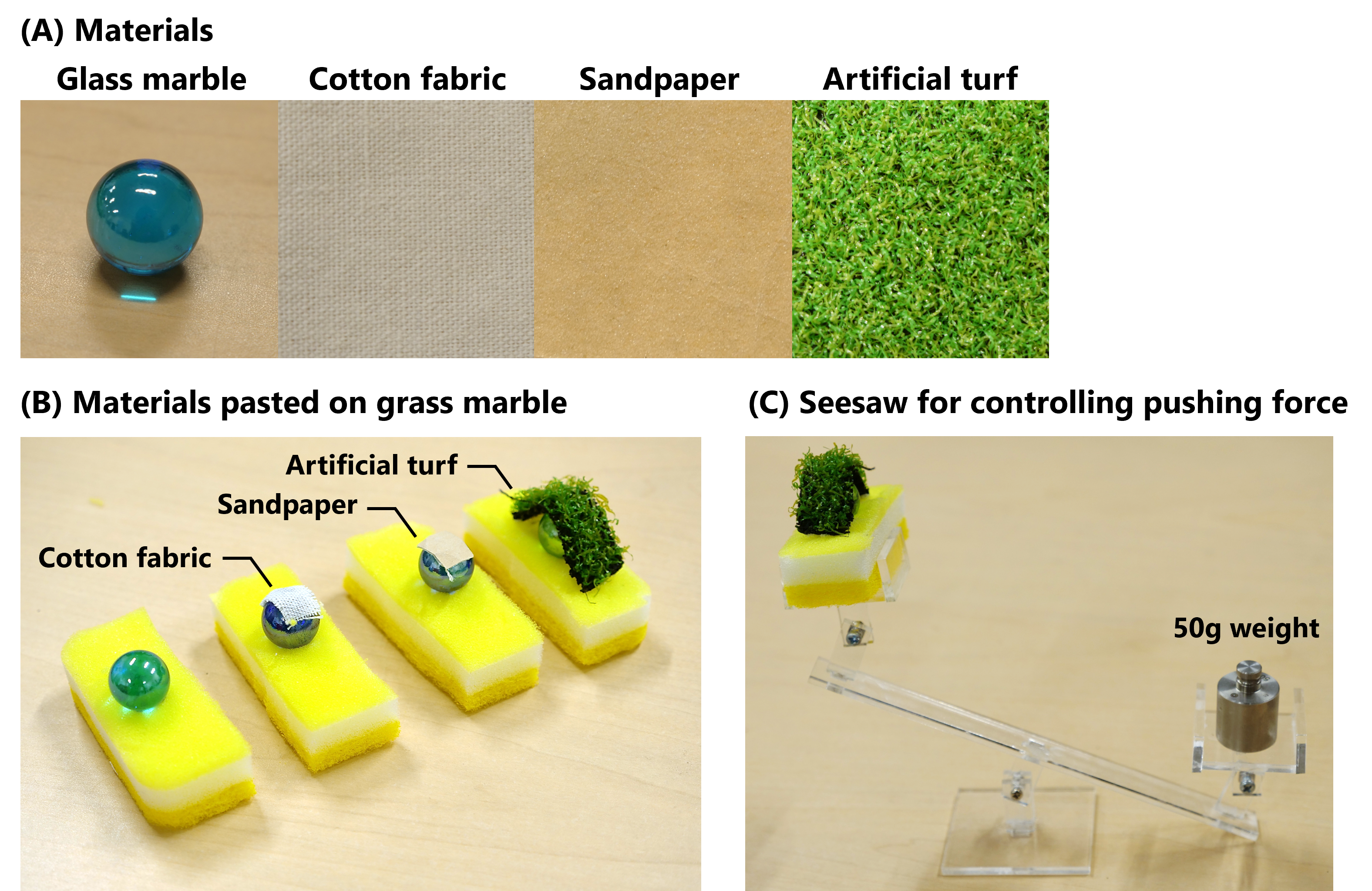}
  \caption{Apparatus used in Experiment 4. (A) Four materials used. (B) Materials pasted on a glass marble. Participants stroked the pasted materials' surface with their palms. (C) An acrylic-made seesaw is used for controlling the pushing force of participants. A counterweight of 50 g was used.}
  \label{fig:Fig/Ex4/Materials.png}
\end{figure}

The setup for this experiment was the same as in Experiment 1 (Fig.~\ref{fig:Fig/Setup/System.png}).

The ultrasound stimuli for this experiment were the same as in Experiment 3. We used six stimuli: S-LM, S-30Hz-w, S-30Hz-s, S-150Hz-w, S-150Hz-s, S-Mix2, as described in Section~\ref{sec:Ex3 Setup and Stimulus}.

Four real materials shown in Fig.~\ref{fig:Fig/Ex4/Materials.png}A were used for comparison with ultrasound stimuli: a glass marble, cotton fabric, sandpaper with a grit size of 100, and artificial turf. These materials were chosen so that each provides a different tactile sensation. For example, a glass marble was chosen to provide the smoothest sensation, while sandpaper was chosen to provide strong roughness and friction sensations. 

The reason for choosing a glass marble was that its shape was close to the tactile shape of an ultrasound stimulus. Morisaki et al. showed that the perceived shape of S-LM is close to a sphere~\cite{morisaki2023noncontact}. Other materials were pasted onto a glass marble so that their tactile shapes were close to that of this ultrasound tactile stimulus (shown in Fig.~\ref{fig:Fig/Ex4/Materials.png}B).

\subsection{Procedure}
\begin{figure*}[t]
  \centering
  \includegraphics[width=1.9\columnwidth]{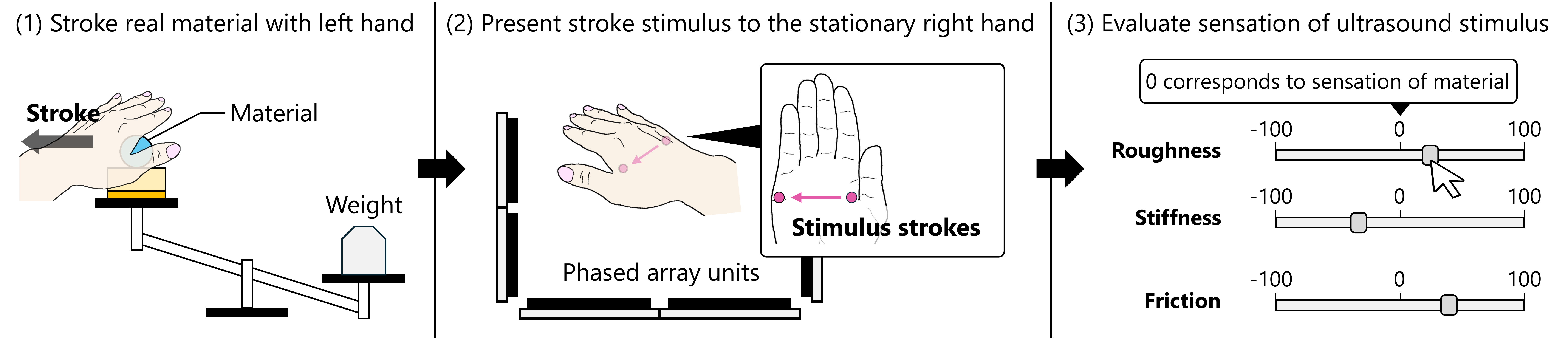}
  \caption{Procedure of Experiment 4. (1) Participant stroked the real material with the left palm. (2) Next, a stroking ultrasound stimulus was presented to the participant’s right palm. The stroking speed was 1.8 cm/s. (3) Participants evaluated the roughness, stiffness, and friction sensations of the experienced ultrasound stimulus using a slider ranging from -100 to 100. A score of 0 corresponded to the same roughness, stiffness, and friction sensations as the real material.}
  \label{fig:Fig/Ex4/Procedure.png}
\end{figure*}
15 males (11 in their 20s and 4 in their 30s) and 4 females (all in their 20s) participated in the experiment. 

Participants experienced the ultrasound tactile stimuli and real materials, and evaluated their sensation of roughness, stiffness, and friction sensations. The procedure was illustrated in Fig.~\ref{fig:Fig/Ex4/Procedure.png}. First, the participant stroked the real material with the left palm. Next, a spherical VR object moved 7 cm in the x-direction across the palm, and the corresponding stroking ultrasound stimulus was presented to the participant’s right palm. The stroking speed of the ultrasound stimulus was 1.8 cm/s. This speed was selected based on preliminary experiments to ensure it felt like a natural stroke. Participants evaluated the roughness, stiffness, and friction sensations of the experienced ultrasound stimulus using a slider ranging from -100 to 100. A score of 0 corresponded to the same roughness, stiffness, and friction sensations as the real material. If participants judged that the ultrasound stimulus was rougher than the touched material, they rated the roughness of the stimulus between 1 and 100 points. Participants adjusted their stroking speed so that it matched the movement of the VR object. In the material stroking, a contact force strength was controlled using a small seesaw shown in Fig.~\ref{fig:Fig/Ex4/Materials.png}C. The real material was placed on this seesaw, and a 50 g weight was placed on the opposite side. Participants controlled their touch strength so that the seesaw did not move. The real materials were replaced manually by the experimenter. Participants could experience the real material and ultrasound stimuli up to three times.

As a control stimulus, a glass marble was used instead of an ultrasound stimulus. In this control condition, participants first stroked the real material as a standard stimulus, then stroked the glass marble as a comparison stimulus. Those strokes were performed with the left hand. After that, they evaluated the roughness, stiffness, and friction sensations of the glass marble on the -100 to 100 scale. In the case of comparison between glass marble, participants were instructed to respond with 0.

The real materials and ultrasound stimuli were presented in a random order. In total, participants experienced 28 times ((6 stimuli + 1 control) $\times$4 materials = 28).

\subsection{Results\label{sec:Ex4 Result}}
\begin{figure*}[t]
  \centering
  \includegraphics[width=2\columnwidth]{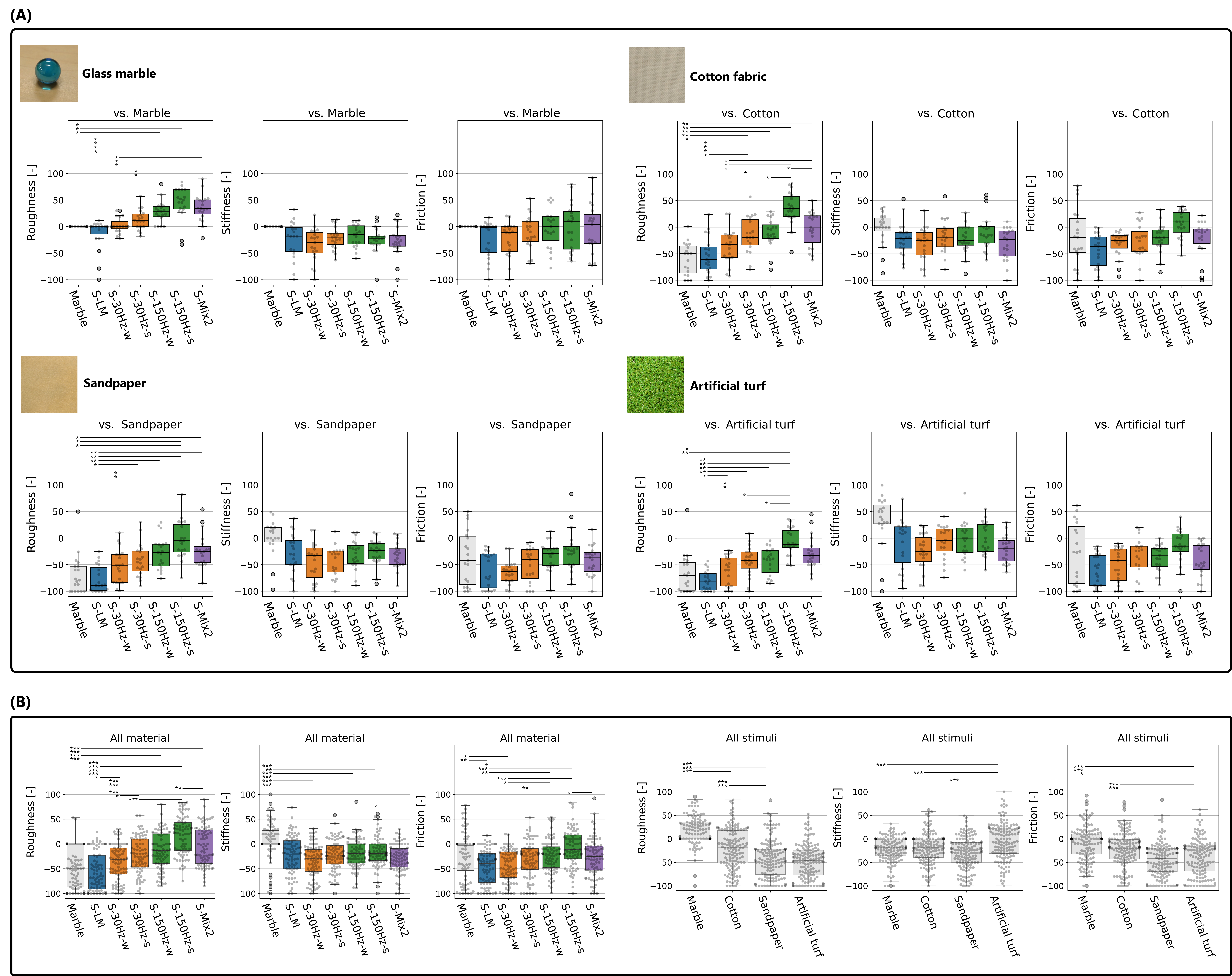}
  \caption{Results of Experiment 4. Comparison results of roughness, stiffness, and friction sensations. (A) Evaluation results for each real material. The results of the pairwise comparison are also shown. This comparison was conducted only on roughness data. *, **, and *** indicate $p < 0.05$, $p < 0.005$, $p < 0.0005$, respectively. (B) The left graph shows the combined data for all materials. The right graph shows the combined data for the stimulus condition. The results of the pairwise comparison are also shown.}
  \label{fig:Fig/Ex4/Texture.png}
\end{figure*}

Fig.~\ref{fig:Fig/Ex4/Texture.png} shows the evaluation results for roughness, stiffness, and friction sensations.

\subsubsection{Analysis of Roughness}
We analyzed the effect of the stimulus condition and the real material on the roughness data. Since a Shapiro-Wilk test showed the roughness data for comparison pairs of [S-LM, marble], [S-LM, sandpaper], and [marble, sandpaper] were not normally distributed ($p < 0.05$), after applying the aligned rank transformation (ART) to the data~\cite{wobbrock2011aligned}, a repeated two-way ANOVA was conducted. The results showed that the stimulus condition ($F(6, 108) = 48.7, p = 1.6\times 10^{-28}, {\eta_p}^{2} = 0.73$), the real material ($F(3, 54) = 74.1, p = 3.9\times 10^{-19}, {\eta_p}^{2} = 0.8$), and their interaction had significant effects ($F(18, 324) = 1.87, p =0.017, {\eta_p}^{2} = 0.094$). As the interaction effect was significant, we examined the simple main effects of stimulus condition and comparison material by applying a Friedman test with Holm correction to the data. The results showed that the stimulus condition had a significant effect on roughness at all real materials ($\chi^2(6), p < 0.0005$), and the real material had a significant effect on roughness at all stimulus conditions ($\chi^2(3), p < 0.0005$).

To examine differences in roughness between stimulus conditions, a Wilcoxon signed-rank test with Holm correction was applied to the roughness data for each material (shown in Fig.~\ref{fig:Fig/Ex4/Texture.png}A). For all materials, the roughness of S-LM was significantly lower than that of all other stimuli except S-30Hz-w and the control. The roughness of S-150Hz-s was significantly higher than that of all other stimuli except S-150Hz-w and S-Mix2. These results are shown in Fig.~\ref{fig:Fig/Ex4/Texture.png}A.

As additional analysis, the multiple comparisons were also conducted on the combined roughness data within materials and stimulus conditions shown in Fig.~\ref{fig:Fig/Ex4/Texture.png}B. These results were also added to each corresponding graph.

These results indicated that the roughness sensation of the ultrasound stimulus was significantly increased by synthesizing vibration sensation.

\subsubsection{Analysis of Stiffness}
We analyzed the stiffness data. Since a Shapiro-Wilk test showed the data for comparison pairs of [marble, artificial turf] was not normally distributed ($p < 0.05$), after applying ART to the data~\cite{wobbrock2011aligned}, a repeated two-way ANOVA was conducted. The results showed that the stimulus condition ($F(6, 108) = 13.5, p = 2.2\times 10^{-11}, {\eta_p}^{2} = 0.43$) and the real material ($F(3, 54) = 10.5, p = 1.6\times 10^{-5}, {\eta_p}^{2} = 0.37$) had significant effects. To examine the differences between stimulus conditions, a Wilcoxon signed-rank test with Holm correction was applied to the stiffness data for all materials shown in Fig.~\ref{fig:Fig/Ex4/Texture.png}B. The stiffness of the glass marble (control stimulus) was significantly higher than that of all ultrasound stimuli ($p < 0.05$). The stiffness of S-150Hz-s was significantly higher than that of S-Mix2.

These results indicated that the stiffness of ultrasound stimuli was softer than that of a glass marble and that it did not vary significantly with respect to the synthesized vibration intensity.

\subsubsection{Analysis of Friction}
We analyzed the friction data. Since a Shapiro-Wilk test showed the data for comparison pairs of [S-Mix2, cotton], [S-LM, sandpaper] were not normally distributed ($p < 0.05$), after applying the ART to the data~\cite{wobbrock2011aligned}, a repeated two-way ANOVA was conducted. The results showed that the stimulus condition ($F(6, 108) = 8.3, p = 2.3\times 10^{-7}, {\eta_p}^{2} = 0.31$) and the real material ($F(3, 54) = 10.3, p = 1.8\times10^{-5}, {\eta_p}^{2} = 0.36$) had significant effects. To examine the differences between stimulus conditions, a Wilcoxon signed-rank test with Holm correction was applied to the data. The friction sensation of S-150Hz-s was significantly higher than that of S-LM, S-30Hz-w, S-30Hz-s, and S-Mix2 ($p < 0.05$). The friction sensation of S-150Hz-w was also significantly higher than that of S-LM and S-30Hz-w.

These results indicated that synthesizing a 150 Hz vibration sensation significantly increased the friction sensation of ultrasound stimuli.

\subsection{Discussion}
\begin{figure}[t]
  \centering
  \includegraphics[width=0.8\columnwidth]{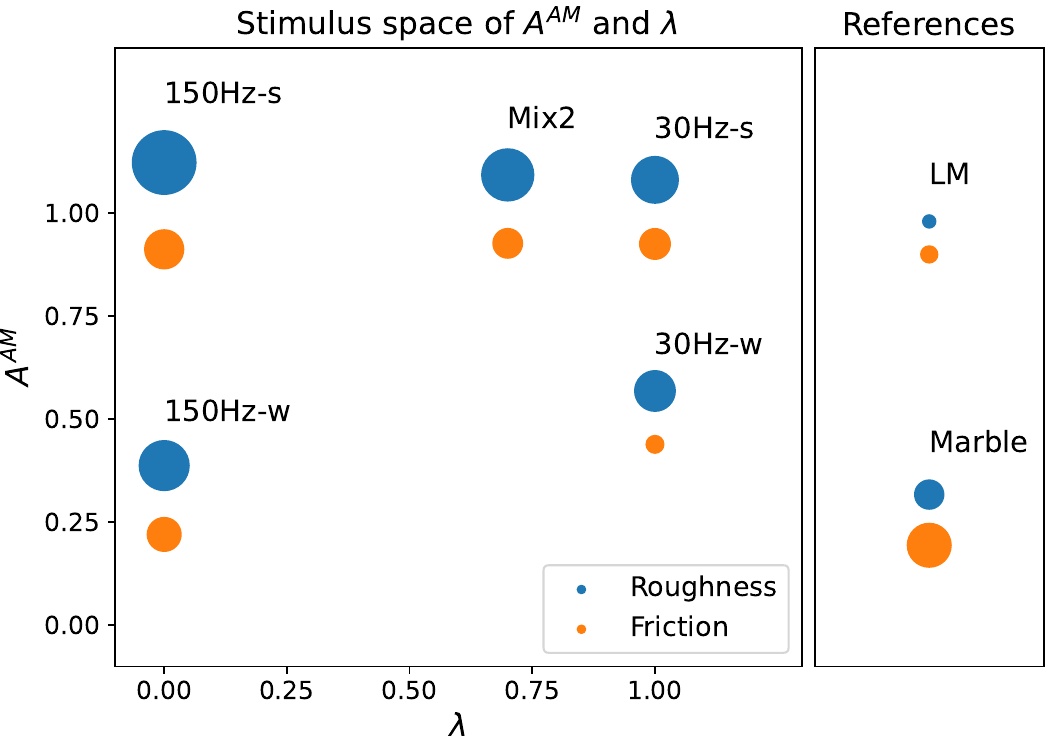}
  \caption{Renderable tactile sensation of the proposed method visualized with $\AmpAM{}$ and $\lambda$. The circle size indicates the median value of roughness and friction data shown in Fig.~\ref{fig:Fig/Ex4/Texture.png}B. S-LM and glass marble were plotted in a separate area because their $\lambda$ was undefined.}
  \label{fig:Fig/Ex4/Bubble.pdf}
\end{figure}

These results demonstrated that pressure sensation evoked by ultrasound can create a smooth and slippery sensation, and its roughness and friction sensations were enhanced by synthesizing vibration sensations while maintaining its stiffness sensations. For enhancing roughness and friction sensations, 150 Hz vibration was more effective than 30 Hz vibration. Only 150 Hz vibration enhanced the friction sensation, while 30 Hz vibrations did not. The stiffness sensation was not varied among most stimulus conditions. To visualize the creatable sensation, the bubble chart of roughness and friction sensations is shown in Fig.~\ref{fig:Fig/Ex4/Bubble.pdf}. The medians were calculated from the data shown in Fig.~\ref{fig:Fig/Ex4/Texture.png}B. The S-LM and glass marble were plotted in a separate area because their $\lambda$ was undefined.

The following section discusses each sensation.

\subsubsection{Roughness}
By synthesizing vibrations, the roughness sensation was significantly enhanced as shown in Fig.~\ref{fig:Fig/Ex4/Texture.png}A and B. This enhancement was consistent with the previous research~\cite{bensmaia2003vibrations, natsume2017skin}. The previous studies showed that the roughness intensity increased as vibration intensity increased~\cite{natsume2017skin}. 

Considering the mechanism of roughness perception, the 30 Hz and 150 Hz vibration may create qualitatively different roughness sensations~\cite{konyo2005tactile,bensmaia2003vibrations}. When stroking a rough surface with different spatial frequencies, vibrations of different frequencies were caused on the fingertips. The frequency of the vibration increased as the surface spatial frequency increased. Therefore, S-30Hz and S-150Hz likely conveyed textures of lower spatial and higher spatial frequency, respectively. 

To examine real materials corresponding to the produced roughness sensations, we applied one-sample Wilcoxon signed-rank tests with Holm correction to the roughness data (shown in Fig.~\ref{fig:Fig/Ex4/Texture.png}A). We confirmed whether the roughness scores were significantly different from zero. There was no significant difference between S-LM and a glass marble, nor between S-150Hz-s and sandpaper ($p > 0.05$).

\subsubsection{Stiffness}
The stiffness sensation of ultrasound stimuli was softer than that of a glass marble, and it was not significantly varied among all stimulus conditions except for [S-150Hz-s, S-Mix2]. The lack of variation in the stiffness sensation may be due to the fact that the maximum sound amplitude $\AmpMax$ was always constant between the stimulus conditions.

Although the stiffness of the ultrasound stimulus was softer than that of a glass marble, the difference was lower than 30 in the median and not huge. Considering that the participants were allowed to apply up to 50 gf of force when stroking, the ultrasound stimulus may not be extremely soft, and it may feel like contact with a physical object. 

The stiffness of the ultrasound stimulus was judged entirely through a passive situation, which may have influenced the results. Ultrasound stimuli were presented to a stationary palm with the same stimulus size (focus size). In contrast, when evaluating the stiffness of real materials, participants actively touched the object and could use changes in contact area between the hand and the materials as a stiffness cue.

We confirmed whether the stiffness scores differed from zero using a Wilcoxon signed-rank test with Holm correction. There was no significant difference in the stiffness between all stimuli except for S-Mix2 and the artificial turf ($p>0.05$).

\subsubsection{Friction}
The friction sensation of the ultrasound stimulus was significantly enhanced by synthesizing a 150 Hz vibration. This friction enhancement was consistent with the perceptual mechanism of friction through vibration. When stroking the surface of an object, depending on its coefficient of friction, the fingertip alternately slips and sticks at several hundred Hz. The frictional vibration at several hundred Hz contributes to the friction sensation~\cite{guest2012physics,nonomura2009tactile}. The synthesized 150 Hz vibration in the ultrasound stimulus may play the role of such frictional vibrations, thereby enhancing the friction sensation. The 30 Hz vibration did not significantly enhance friction sensation, likely because its frequency order of 30 Hz differed from that of typical frictional vibration.

Finally, we confirmed whether the friction scores differed from zero using a Wilcoxon signed-rank test with Holm correction. There was no significant difference in the friction between S-150Hz-s and [glass marble, cotton fabric, and artificial turf] ($p>0.05$).

\section{Limitation and Future Work}
\subsection{Displayable Sensation with Proposed Method}
With the proposed rendering method, reproducing the tactile sensation of objects with high friction and low roughness is difficult. Synthesizing a 150 Hz vibration was necessary to enhance friction sensation, but in this case, roughness sensation was also significantly enhanced. We will explore a method to selectively enhance the sensation of friction.

The perception of stiffness was limited to the level of artificial turf with the proposed method. As a possible solution for this, previous studies have proposed enhancing the stiffness sensation in ultrasound tactile stimulus by dynamically varying the stimulus presentation area according to finger movement~\cite{sun2024expanding}. We will combine this stimulus area variation method with the proposed method to expand the range of displayable stiffness sensations.

Our approach excluded SA-II (slowly adaptive type II) components. Although the SA-II receptor primarily reacts to shear force stimulus, current ultrasound midair haptics cannot cover such stimulus~\cite{johansson1983tactile}.

\subsection{Experimental Design}
In Experiment 4, participants actively stroked real materials, but an ultrasound stimulus was presented to stationary participants' hands, corresponding to a passive touch situation. This difference in the touch situation may have influenced the evaluation results. We will evaluate this effect. 

In Experiment 4, the frequency of the vibration stimulus and the stimulus movement speed was constant. Since roughness perception relies on the vibration frequency perceived when stroking an object surface, changing this frequency and the movement speed may alter the roughness sensation~\cite{konyo2005tactile,bensmaia2003vibrations}. We will evaluate these effects in the future.

\section{Conclusion}
This study proposes a midair tactile rendering method synthesizing three basic ultrasound stimuli: static pressure, 30 Hz vibration, and 150 Hz vibration, primarily stimulating SA-I, FA-I, and FA-II receptors, respectively. The pressure stimulus was presented with a 5 Hz ultrasound foci rotation, while vibration stimuli were synthesized by modulating the sound amplitude of the foci. 

The proposed method can render at least six discriminable tactile textures with different roughness and friction sensations. The pressure-only stimulus felt smooth and slippery, and its smoothness was close to that of a glass marble. The synthesized vibration enhanced its roughness and friction sensations, and its roughness reached that of a 100-grit sandpaper.


%

\ifCLASSOPTIONcaptionsoff
  \newpage
\fi



\bibliographystyle{IEEEtran}
\bibliography{ref}

%


\begin{IEEEbiography}[{\includegraphics[width=1in,height=1.25in,clip,keepaspectratio]{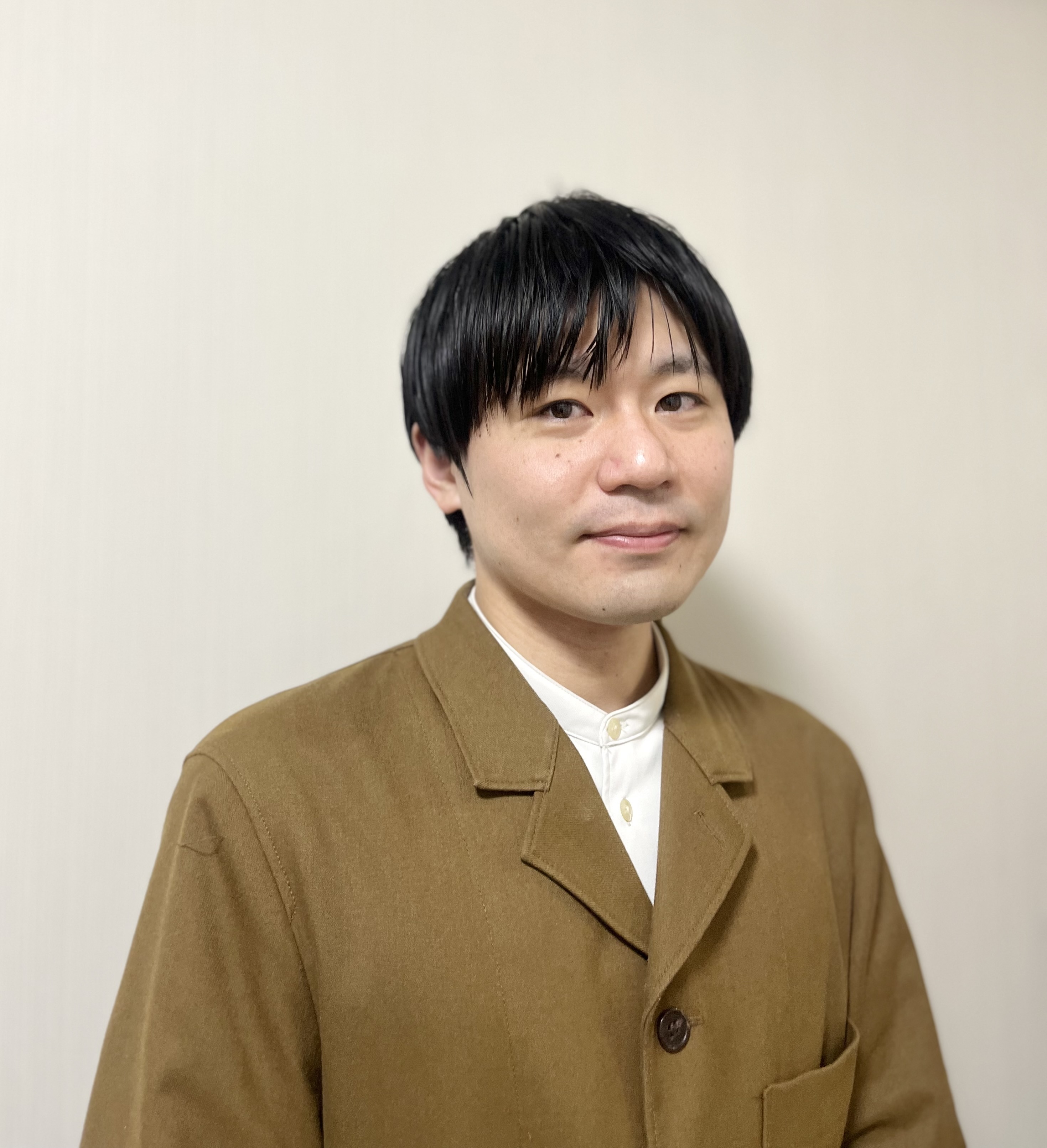}}]{Tao Morisaki}
Tao Morisaki is a researcher at NTT Communication Science Laboratories. His previous affiliation was The University of Tokyo, Japan. He received the M.S. degree in 2020 and the Ph.D. degree in 2023 from the Department of Complexity Science and Engineering, The University of Tokyo. His research interests include haptics, ultrasound midair haptics, and human-computer interaction. He is a member of IEEE and VRSJ.
\end{IEEEbiography}
\begin{IEEEbiography}[{\includegraphics[width=1in,height=1.25in,clip,keepaspectratio]{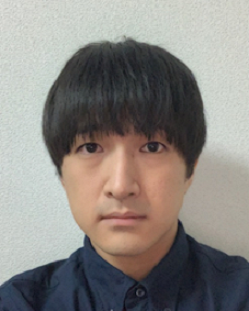}}]{Atsushi Matsubayashi}
Atsushi Matsubayashi received the B.S. degree in engineering and the M.S. and Ph.D. degrees in information science and technology from The University of Tokyo, Japan, in 2015, 2017, and 2020, respectively. He is currently a Project Assistant Professor with the Graduate School of Frontier Sciences, The University of Tokyo. His research interests focus on mid-air haptics.
\end{IEEEbiography}
\begin{IEEEbiography}[{\includegraphics[width=1in,height=1.25in,clip,keepaspectratio]{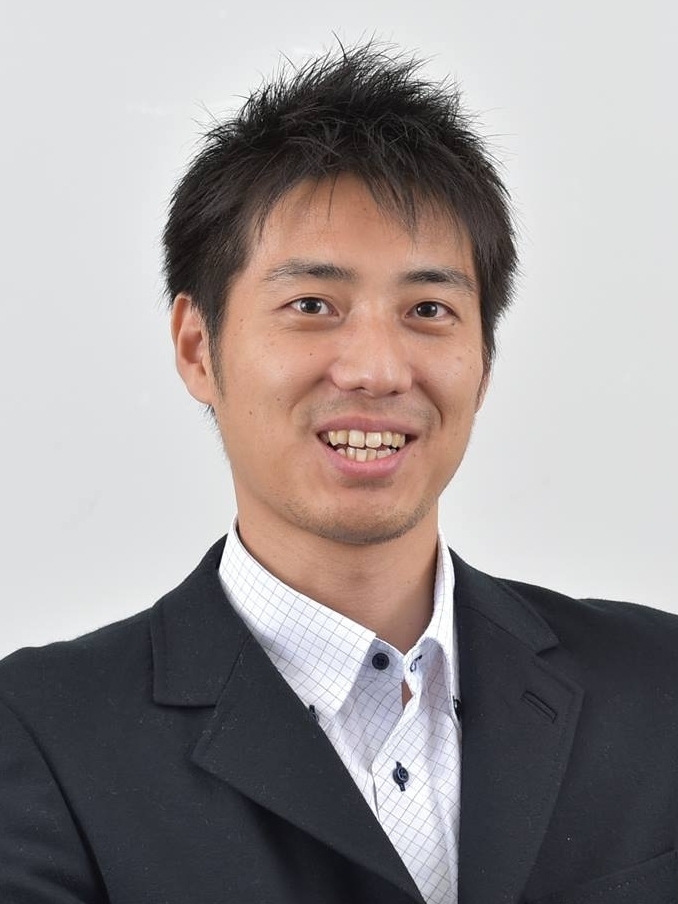}}]{Yasutoshi Makino}
Yasutoshi Makino is an associate professor in the Department of Complexity Science and Engineering in the University of Tokyo. He received his PhD in Information Science and Technology from the Univ. of Tokyo in 2007. He worked as a researcher for two years in the Univ. of Tokyo and an assistant professor in Keio University from 2009 to 2013. From 2013 he moved to the Univ. of Tokyo as a lecture, and he is an associate professor from 2017. His research interest includes haptic interactive systems.
\end{IEEEbiography}
\begin{IEEEbiography}[{\includegraphics[width=1in,height=1.25in,clip,keepaspectratio]{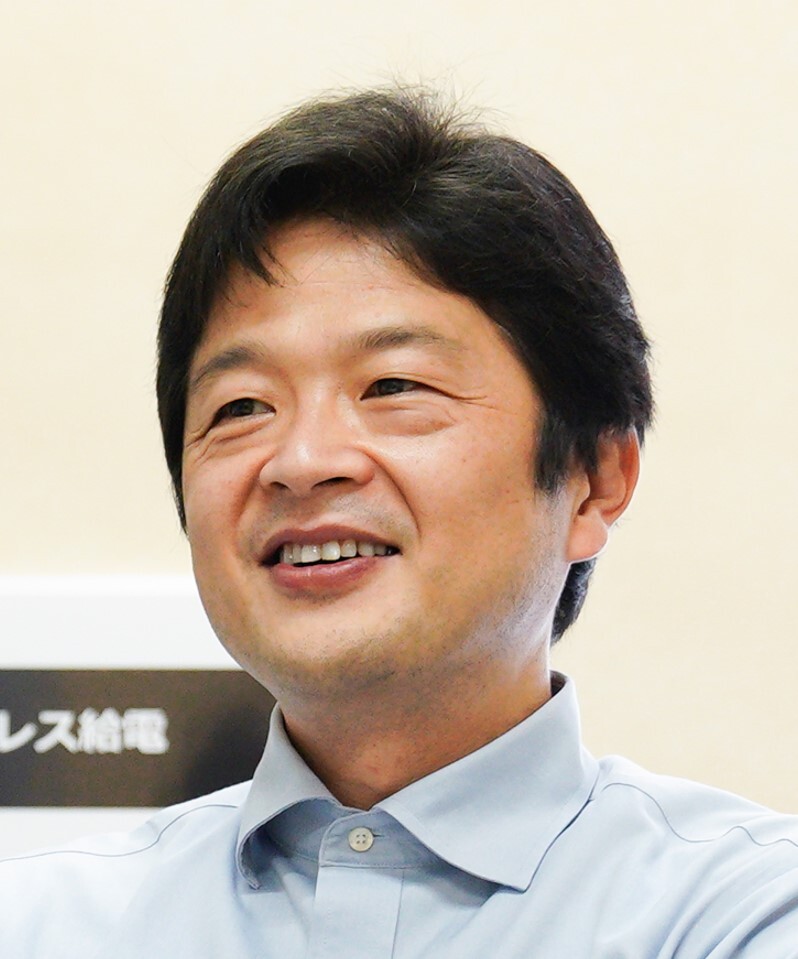}}]{Hiroyuki Shinoda}
Hiroyuki Shinoda is a Professor at the Graduate School of Frontier Sciences, the University of Tokyo. After receiving a Ph.D. in engineering from the University of Tokyo, he was an Associate Professor at Tokyo University of Agriculture and Technology from 1995 to 1999. He was a Visiting Scholar at UC Berkeley in 1999 and was an Associate Professor at the University of Tokyo from 2000 to 2012. His research interests include information physics, haptics, mid-air haptics, two-dimensional communication, and their application systems. He is a member of SICE, IEEJ, RSJ, JSME, VRSJ, IEEE and ACM.
\end{IEEEbiography} 

\vfill


\end{document}